\documentclass[jgrga]{AGUTeX}

\usepackage{graphicx}

\setkeys{Gin}{draft=false}

\authorrunninghead{Raouafi et al.}

\titlerunninghead{Evanescent Clumps from Comet C/2011~L4}
\authoraddr{Corresponding author: N.-E. Raouafi, The Johns Hopkins University Applied Physics Laboratory, Laurel, MD, USA. (NourEddine.Raouafi@jhuapl.edu)}
\begin{document}
\title{Dynamics of High-Velocity Evanescent Clumps [HVECs] Emitted from Comet C/2011~L4 as Observed by STEREO}

\authors{N.-E. Raouafi\altaffilmark{1}, C. M. Lisse\altaffilmark{1}, G. Stenborg\altaffilmark{2},
G. H. Jones\altaffilmark{3}, and C. A. Schmidt\altaffilmark{4}}
\altaffiltext{1}{The Johns Hopkins University Applied Physics Laboratory, Laurel, MD, USA.}
\altaffiltext{2}{SPACS, College of Science, George Mason University, Fairfax, VA, USA.}
\altaffiltext{3}{UCL Mullard Space Science Laboratory, Dorking, Surrey, UK.}
\altaffiltext{4}{Materials Science \& Engineering, University of Virginia, Charlottesville, VA, USA.}

\begin{abstract}
High-quality white-light images from the SECCHI/HI-1 telescope onboard STEREO-B reveal high-velocity evanescent clumps [HVECs] expelled from the coma of the C/2011~L4 [Pan-STARRS] comet. The observations were recorded around the comet's perihelion [i.e., $\sim~0.3$~AU] during the period $09-16$ March 2013. Animated images provide evidence of highly dynamic ejecta moving near-radially in the anti-sunward direction. The bulk speed of the clumps at their initial detection in the HI1-B images range from  $200-400$~km~s$^{-1}$ followed by an appreciable acceleration up to speeds of $450-600$~km~s$^{-1}$, which are typical of slow to intermediate solar wind speeds. The clump velocities do not exceed these limiting values and seem to reach a plateau. The images also show that the clumps do not expand as they propagate. The while-light images do not provide direct insight into the composition of the expelled clumps, which could potentially be composed of fine, sub-micron dust particles, neutral atoms and molecules, and/or ionized atomic/molecular cometary species. Although solar radiation pressure plays a role in accelerating and size sorting of small dust grains, it cannot accelerate them to velocities $>200$~km~s$^{-1}$ in the observed time interval of a few hours and distance of $<\!10^6$~km. Further, order of magnitude calculations show that ionized single atoms or molecules accelerate too quickly compared to observations, while dust grains micron sized or larger accelerate too slowly. We find that neutral Na, Li, K, or Ca atoms with $\beta>50$ could possibly fit the observations. Just as likely, we find that an interaction with the solar wind and the heliospheric magnetic field (HMF) can cause the observed clump dynamical evolution, accelerating them quickly up to solar wind velocities. We thus speculate that the HVECs are composed of charged particles (dust particles) or neutral atoms accelerated by radiation pressure at $\beta>50$ values. In addition, the data suggest that clump ejecta initially move along near-radial, bright structures, which then separate into HVECs and larger dust grains that steadily bend backwards relative to the comet's orbital motion due to the effects of solar radiation and gravity. These structures gradually form new striae in the dust tail. The near-periodic spacing of the striae may be indicative of outgassing activity modulation due to the comet nucleus' rotation. It is, however, unclear whether all striae are formed as a result of this process.
\end{abstract}

\begin{article}
\section{Introduction}
Although comets have been observed for centuries, phenomena related to their morphologies, composition, formation of neutral, ion, and dust tails, and interaction with the local environment [e.g., solar wind] are still not completely understood. Observations show that cometary nuclei, which are composed of dusty ices [i.e., water and other volatile ices, such as CO, CO$_2$], have fragile structures that are susceptible to heat-driven fragmentation \citep[e.g., Shedding or sloughing, fission or shattering, disintegration; see][]{2004JGRE..10912S03C}, sublimation, evaporation, and photoionization of diverse molecular species such as H$_2$O, CO, N$_2$, CO$_2$, OH, CN, CH, NH, C$_3$, NH$_2$ \citep[][]{1968ARA&A...6..267B,2014Icar..239..141L}.

Early ground-based observations of comets led to the conclusion that cometary activity must originate from discrete locations on the nuclei surfaces. This was based on the diverse morphologies within cometary comas such as jets, expanding halos, and fan-shaped comas. These are thought to be the result of the combined effects of the localized activity on the nuclei surfaces and their rotation as well as their Sun-comet [seasonal variations] and Earth-comet geometries \citep[see][and references therein]{2007SPIE.6694E..0IS}. Close-up images of comet nuclei from flyby space missions [e.g., Giotto: comets Halley and Grigg-Skjellerup, \citealt[][]{1986Natur.321..320K,1987A&A...187..807K}; Deep Space 1: comet 19P/Borrelly, \citealt[][]{2002Sci...296.1087S,2002LPI....33.1256S}; Stardust: comet 81P/Wild, \citealt[][]{2004Sci...304.1764B}; Deep Impact: comet Tempel 1, \citealt[][]{2005Sci...310..258A}; and Rosetta: comet 67P/Churyumov-Gerasimenko, \citealt[][]{2007SSRv..128....1G}] could only confirm this conclusion \citep{2004Sci...304.1764B,2004Sci...304.1769S}. They particularly show that most of the nuclei surfaces are dark, inert, inactive, and refractory. The heat-driven activity of the cometary nuclei is contained in isolated bright sites on the surface or from subsurface cavities \citep{2004Icar..167...30Y} where gases escape mostly sunward in the form of jets carrying with them chunks of ice-mixed dust as proposed by \cite{1951ApJ...113..464W} in his conglomerate model \citep[see also][]{2004JGRE..10912S03C}.  

Dust-tail striae are thought to form through the fragmentation of dust agglomerates \citep{1980AJ.....85.1538S,1982AJ.....87.1836S,1980ApJ...242..395H,1988AN....309..133N,1990Icar...87..403N,1997EM&P...78..329P}. The escaping gases drag ice-mixed dust forming a strongly coupled ensemble in the vicinity of the nucleus, which then decouple as they expand further out leaving the several-hundreds m~s$^{-1}$ fast icy conglomerates exposed to the solar conditions \citep[solar heat, electrostatic charging, collisions, etc.; see][]{1990LPI....21..653K,2004JGRE..10912S03C}. The exposure of the ejected material leads to further sublimation of the ices, which causes further fragmentation of bare, porous, and fragile dust aggregates. Radiation pressure forces gradually modify the dust and gas trajectories anti-sunward and then into the different tails [i.e., dust, ion, and neutral tails]. Another discovery from the Stardust mission supporting the theories of dust aggregate fragmentation was reported by \cite{2004Sci...304.1776T}. They found strong fluctuations in particle number density in 100 milliseconds or less corresponding to distances $\ll1$~km, which at times were intercepted by gaps with very low to practically no particles \citep[see also][]{2004JGRE..10912S03C}. For a review  on the dynamics of cometary comas, see \cite{2007SPIE.6694E..0IS}.


The pick up of the ionized species by the solar wind and the embedded heliospheric magnetic field [HMF] results in the formation of the plasma [or ion] tail, which is susceptible to electromagnetic forces of the ambient environment and changes in the solar wind conditions. The plasma tail is hence swept anti-sunward with a velocity-aberration correction of a few degrees depending on the comet's speed relative to that of the solar wind \citep{1976P&SS...24..287E,1986GeoRL..13..239M,1982come.coll..561R,1994ApJS...91..419F,2004DPS....36.2102J,2008ApJ...677..798B,2010ApJ...713..394C,2013AIPC.1539..364J}. Along with the volatile species, dust particles with different sizes lift-off from the nucleus surface into the cometary coma where they become exposed to an array of physical processes \citep[e.g., gravity, radiation pressure, fragmentation, charging, collisions, etc.;][]{1993Icar..101...84K,2002Icar..157..349K,2013RvGeo..51...53M} dictating their subsequent dynamical evolution. 

Large dust particles [i.e., $>1$~$\mu{\rm{m}}$] are typically dominated by the effects of solar gravity and radiation pressure forces and form a rather morphologically diffuse tail. Thus, the dust tail, which is formed by particles that are individually in orbit around the Sun, varies from linear anti-solar to curved along the orbit \citep{1998ApJ...496..971L,2004Icar..171..444L}. The range of ratios of solar gravity to solar radiation forces on these particles causes the size sorting as well as the dust tail morphology to curve as the comet swings around the Sun. The orbital motions of supra-micron particles may only be affected by extreme solar wind events, such as coronal mass ejections \citep[CMEs;][]{2004GeoRL..3120805J,2009ApJ...696L..56J}. Due to solar photon driven photoelectric effect and solar wind sputtering, it is also expected that dust grains become charged and every low mass particle [$<0.1\ \mu{\rm{m}}$] becomes susceptible to the effects of the HMF and the solar wind \citep[e.g.,][]{1991GMS....61...17M,2014Icar..236..136K}.

\begin{figure*}[!th]
\begin{center}
\includegraphics[width=0.9\textwidth]{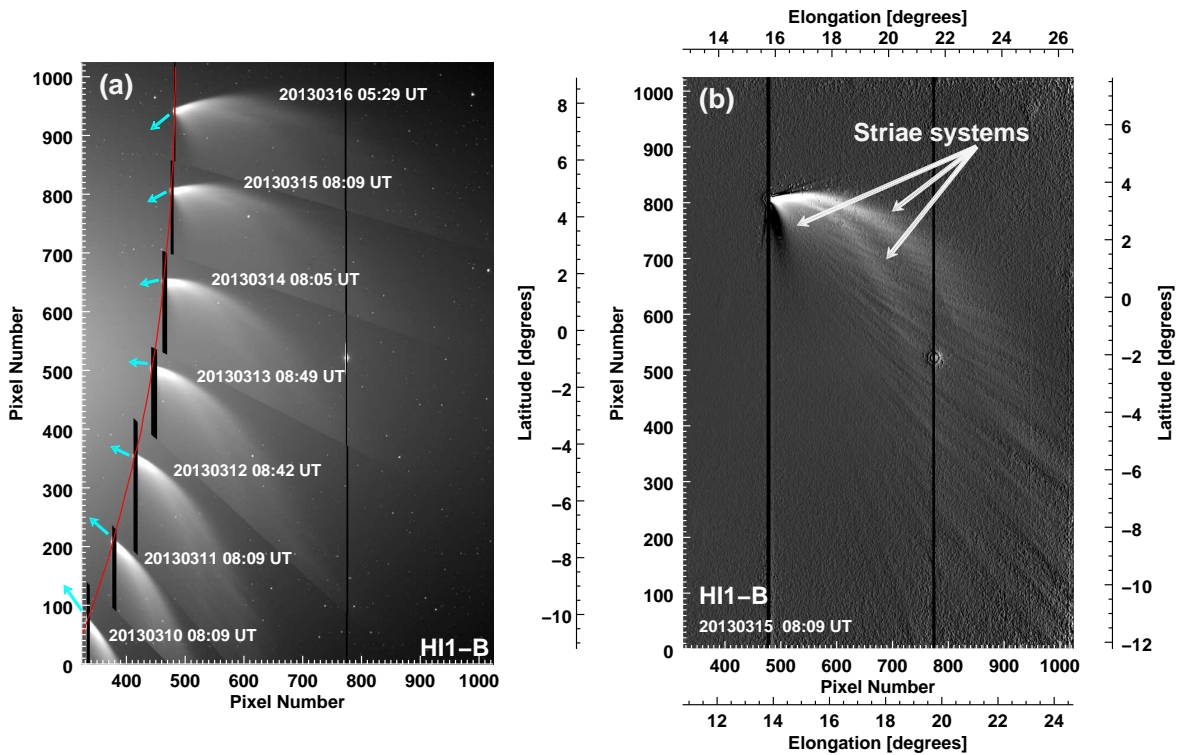}
\caption{(a) Composite intensity images of STEREO/SECCHI/HI-1B showing different phases of the C/2011~L4 comet. The recording dates of the different images are indicated. The red curve show the orbit of the comet and the arrows point toward Sun center. (b) Filtered image  highlighting the rate of local intensity changes. The dust tail is particularly noticeable with at least three systems of striae, which are pointed to by arrows. The dark vertical lines are saturated pixel arrays due to CCD bleeding.}
\label{FIGURE_1}
\end{center}
\end{figure*}

C/2011~L4 [Pan-STARRS; Figure~\ref{FIGURE_1}], a non-periodic comet, was discovered on June 6$^{th}$, 2011 at 7.9 AU heliocentric distance with an apparent magnitude of 19 \citep{2011IAUC.9215....1W}. It was first observed by the 1.8~m Pan-STARRS~1 survey telescope [Haleakala, HI] and  later by the CFHT [http://www.cfht.hawaii.edu]. It became visible to the naked eye, with a visual magnitude of -1.0, near its perihelion at ~0.3 AU in February-March 2013. It is one of the brightest comets since the great comet C/1995~O1 \citep[Hale-Bopp;][]{2014ApJ...784L..23Y}. Orbital analyses [Figure~\ref{FIGURE_2_new}] indicate that it is likely a new comet from the Oort cloud \citep[e.g.,][]{2012A&A...544A.119K}. C/2011~L4 provide therefore a great opportunity to monitor the composition and dynamics of relatively pristine, unprocessed comets.

The dust and gas production of comet C/2011~L4 is enigmatic. The comet was more active before perihelion than after \citep{2014AJ....147..126C}. C/2011~L4 is an unusually dust-rich comet with very low gas emission \citep[][]{2013ApJ...771L..21F,2014ApJ...784L..23Y}. The comet activity was first observed beyond the water-ice sublimation zone \citep[i.e., $5-6$~AU from the Sun;][]{2014ApJ...784L..23Y}. This early activity may be explained by latent heat release from the amorphous-crystalline water ice transition; sublimation of frozen super-volatiles; and/or comet fragmentation \citep{1992ApJ...388..196P,2003Icar..162..183N,2003Icar..161..157B,2004come.book..317M,2004come.book..301B,2014ApJ...784L..23Y}. Spectral observations of C/2011~L4 showed strong emission in the 2.0~$\mu$m spectral band and a rather weak emission in the 1.5~$\mu$m band \citep{2014ApJ...784L..23Y}. \citet{2014ApJ...784L..23Y} showed through Mie modeling that the ratio between the depths of these spectral bands is sensitive to ice grain size. They particularly found that the inclusion of sub-micron grains allows the prominent 2.0~$\mu$m and the much weaker 1.5~$\mu$m bands to be reproduced fairly well. They argue that the comet spectra may be explained by $\sim30\%$ very fine-grained [i.e., $\sim0.2$~$\mu$m] ice mixed with spectrally featureless materials and that the distant activity of C/2011~L4 could be driven by highly volatile ices such as CO$_2$.

C/2011~L4 was observed by the Sun-Earth Connection Coronal and Heliospheric Investigation \citep[SECCHI:][]{2008SSRv..136...67H} heliospheric imagers HI-1 and HI-2 \citep{2000SPIE.4139..284S} onboard the Solar TErrestrial RElations Observatory \citep[STEREO:][]{2008SSRv..136....5K} Behind [i.e., STEREO-B] on March 09--16, 2013. The high quality of the recorded images show evidence for abundant dust in the comet's coma and tail with rich dynamical structures, including multiple straie [Figure~\ref{FIGURE_1}] as well as a series of unusual high-velocity evanescent clumps [hereafter HVECs] propagating anti-sunward [Figure~\ref{FIGURE_2}]. 

The main aim of this work is the characterization the HVECs and their interaction with the local solar wind as well as their connection to newly formed striae in the dust tail based mainly on STEREO/HI-1B observations. The paper is organized as follows: observations and data processing are briefly described in Sect. 2. Sect. 3 is dedicated to the data analysis. Discussion of the results and conclusions are given in Sect. 4. 

Although we speculate on the processes at the origin of the HVECs and physical mechanisms that may influence their dynamical behavior, rigorous modeling of dynamics of these features is out of the scope of this paper and is therefore left for a future publication.

\section{Observations and Data Processing}
The STEREO payload comprises two white-light imagers \citep[termed HI-1 and HI-2 with respective spectral bandpasses of $630-730$~nm $400-1000$~nm and near-flat quantum efficiency of the detectors of $>90\%$ within the wavelength range $500-700$~nm; see][]{2009SoPh..254..387E}. They are dedicated to heliospheric observations with combined field of views [FOVs] covering solar elongations ranging from $4.0^\circ$ to $88.7^\circ$. The HI-1 and HI-2 optical axes are nominally set in the ecliptic plane at $14.0^\circ$ and $53.7^\circ$ from the Sun, with respective FOVs of $20^\circ$ and $70^\circ$. The detectors are $2048\times2048$ arrays with $13.5\times13.5\ \mu{\rm{m}}^2$ pixel size. The onboard $2\times2$ binning results in respective image angular resolutions of $\sim\!105.0^{\prime\prime}$ and $\sim\!6.0^{\prime}$. Details of the optical specifications and calibration of the HI-1 and HI-2 instruments are given by \citet{2009SoPh..254..387E} and \citet{2012SoPh..276..491B}.

\begin{figure*}[!t]
\begin{center}
\includegraphics[width=0.87\textwidth]{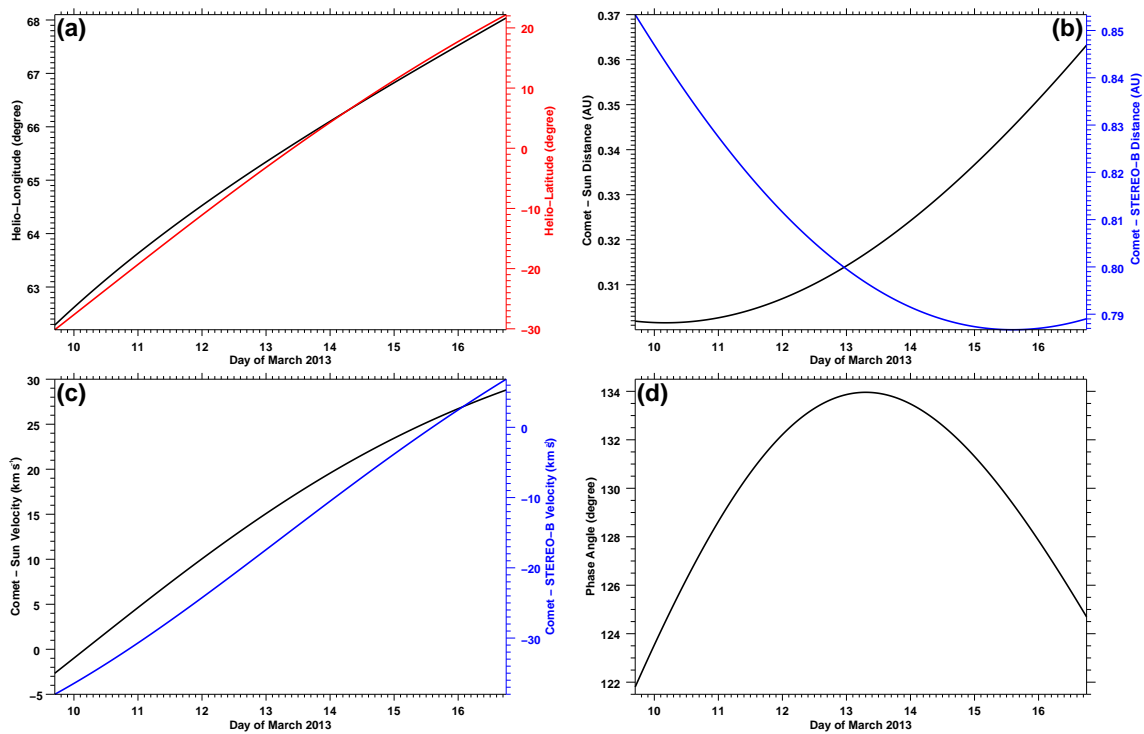}
\caption{C/2011~L4 orbital parameters during the period of the crossing the field-of-view of STEREO/HI1-B. (a) Helio-longitude [black] and helio-latitude [red]; (b) Comet heliodistance [black] and distance to STEREO-B [blue]; (c) Velocity of the comet with respect to the Sun [black] and STEREO-B [blue]. Negative [positive] velocities mean the comet is moving toward [away from] the Sun or STEREO-B. (d) Phase angle of the comet. Data source:  http://ssd.jpl.nasa.gov/horizons.cgi.}
\label{FIGURE_2_new}
\end{center}
\end{figure*}

C/2011~L4 crossed the HI-1B FOV between March 9$^{th}$, 2013 $\sim\!\!$~16:50~UT and March 16$^{th}$, 2013 $\sim\!\!$~18:00~UT  [see Figure~\ref{FIGURE_1}]. Figure~\ref{FIGURE_2_new} provides the orbital parameters of the comet within the HI1-B FOV for the period March 9-16, 2013. The phase angle (Sun-C/2011~L4-STEREO B) varied from $\sim\!122^\circ$ to $\sim\!134^\circ$. This section of the comet orbit was mostly South-North oriented with a nearly constant speed of $\sim\!95$~km~s$^{-1}$ as seen from STEREO-B [i.e., projected on the plane of the sky]. With respect to the Sun, the comet's velocity varies from $-2.7$~km~s$^{-1}$ to 28.8~km~s$^{-1}$. The comet reached its perihelion distance of $\sim\!0.301$ AU on March $10^{th}$ at 04:04~UT \citep{2014AJ....147..126C}. As seen from STEREO-B, the comet's apparent closest distance to the Sun was $\sim\!49.0$~${\rm{R}}_\odot$ [i.e., 0.23 AU] on March $12^{th}$ at 18:50~UT. The extent of the dust tail, which was observed with an angle $\sim\!130^\circ$ between the line of sight and the comet's orbit plane by both HI1-B and  HI-2B, was greater than $10^7$~km [$\sim\!100$~R$_\odot$].  It remained visible in HI-1B and HI-2B FOVs until March $19^{th}$ and $22^{nd}$, respectively.

The present analysis is mainly based on white-light images recorded by STEREO/SECCHI/HI-1B during the period of March 09-16, 2013. The images are recorded with a cadence of 40 minutes. Images from HI-2 [not shown here] recorded with a cadence of 2 hours were used for qualitative analysis. Level 0.5 data acquired from the Virtual Solar Observatory [VSO: http://www.virtualsolar.org] database are processed to higher levels using the instruments' calibration procedures available in the Solar Software tree [SSW: http://www.lmsal.com/solarsoft]. The STEREO observations provide unprecedented details of the morphology of the dust tail, including multiple stria systems [see Figure~\ref{FIGURE_1}b]. A novel phenomenon of HVECs presumably ejected from the comet nucleus is also clearly noticeable in running-difference images as they propagate anti-sunward [see Figure~\ref{FIGURE_2}]. 

The focus in this work is on these transient bright clumps; we leave dynamical analysis of the comet's large dust particle tail to a future paper. Based on the images of HI1-B we determine the clump's dynamical properties as well as their relationship with newly forming striae in the dust tail. The use of the SSW procedures alone was insufficient to analyze the bright clumps and supplemental image processing was necessary. In particular, we used techniques such as homomorphic filtering \citep{OppenheimEtAl1968}, to simultaneously normalize the brightness and increase the contrast across the HI images, and hence make clearer the faint transient features of interest and the fine structure within the dust tail. As implemented, the procedure helps remove the effects of sky background [i.e., light scattered off electrons and dust: K- and F-corona, respectively]. The star field was minimized by applying a sigma filter available in the SSW. For example, compare the raw intensity image in Figure~\ref{FIGURE_1}a and the processed image in Figure~\ref{FIGURE_1}b.

\section{Data Analysis}
Table~\ref{TABLE_1} lists 12 HVECs expelled from the C/2011~L4 coma. The dates and times of their appearance, the corresponding heliocentric and cometocentric distances, position angles, and dates and times of their disappearance are also shown. The sizes of these structures are determined using the running-difference images. The values shown in Table~\ref{TABLE_1} are the averages of the measured size at different times, which do not change significantly. The size error is the standard deviation of the measurements divided by their numbers. The running-difference images displayed in Figure~\ref{FIGURE_2} illustrates several of the HVECs. A number of these clumps could also be traced within the HI-2B FOV, particularly the ones ejected northward of the ecliptic plane. It is, however,  harder to characterize their propagation within HI-2B FOV because of the lower cadence and resolution of the HI-2B observations. It is clear from Figure~\ref{FIGURE_2} and Table~\ref{TABLE_1} that the number of ejecta increased after the comet crossed the ecliptic plane, particularly on March 14--15. The dynamical evolution of the ejected clumps as observed by HI1-B is shown in Figure~\ref{FIGURE_3}.

Figure~\ref{FIGURE_3}a,b show the propagation trajectories of the HVECs as they crossed the HI1-B FOV anti-sunward. The clump appearance correspond to the moment when a coherent structure is distinguishable from the other structures and can be followed for several frames. Their  disappearance is the moment when they can no longer be clearly discernible from background. The clump positions correspond to their approximate centers as seen in the running-difference images. They are determined interactively by hand using remapping of the images from pixel elongation into heliocentric distance and position angle. The remapping software is included in the STEREO package in the SSW tree. The error on the HVECs' positions is assumed to be equal to their respective size precision, which is a conservative approach. They follow nearly radial lines where the change of the position angle is $<2^\circ$ in all cases [Figure~\ref{FIGURE_3}b]. This is evidence for near-radial propagation of the HVECs with respect to the Sun. The ejecta propagation paths may bear clues to their interaction with the ambient solar wind that is presumably propagating radially in this region of the Solar System.

Since the excess in white light emission can be due to scattering by free electrons and/or dust grains, the white light data do not directly provide insights into the composition of these clumps. These may be composed of sub-micron dust particles that are swept away under the influence of solar radiation pressure, the solar wind, and charging effects. They may also be composed of neutral or ionized atomic/molecular species \citep[e.g., Na$^{+0}$ or K$^{+0}$, Li$^{+0}$ or Ca$^{+0}$ or Na$^+$ or K$^+$ alkali ions; see][]{2007ApJ...661L..93F,2013ApJ...771L..21F}.

\subsection{Clump Dynamics}
\begin{figure}[!h]
\begin{center}
\includegraphics[width=0.4\textwidth]{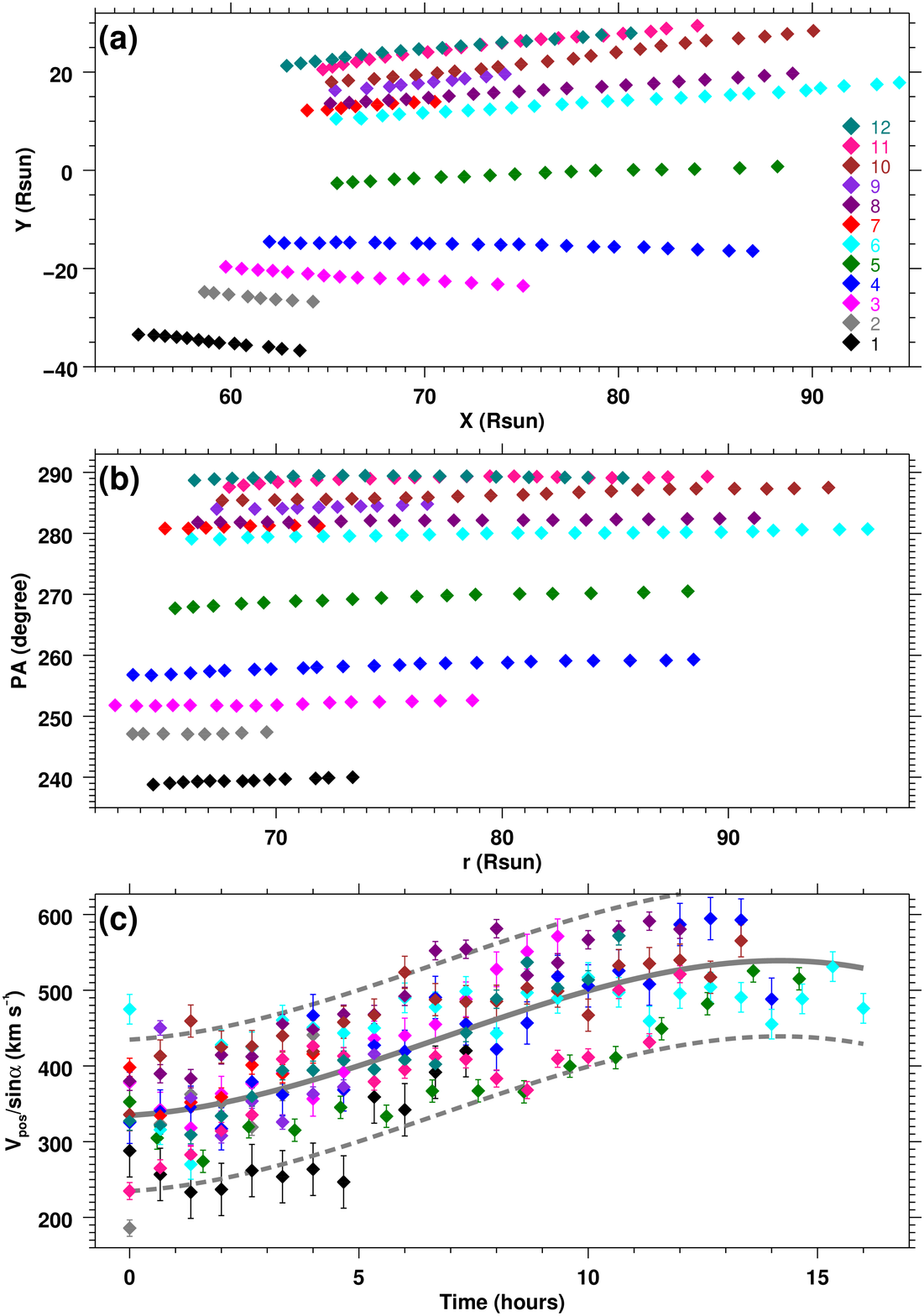}
\caption{(a-b) Propagation paths of the 12 HVECs listed in Table~\ref{TABLE_1}. The symbols and the corresponding clump numbers of Table~\ref{TABLE_1} are shown in the top panel. The clump trajectories are near-radially aligned as illustrated by the position angles that vary by less than $2^\circ$ for all clumps. (c) Bulk velocity of the HVECs corrected for projection effects [i.e., $V_{pos}/\sin\alpha$, where $V_{pos}$ is the plane-of-the-sky speeds of the clumps as estimated from the observed motion of clumps in the HI1-B images. $\alpha$ is the phase angle of the comet]. The solid grey curve is the best polynomial fit of the clump propagation velocities. The dashed lines show that more than 90\% of measured velocities lie within the best curve $\pm100\ {\rm{km\ s}}^{-1}$. The HVECs are accelerated mainly in the radial direction from initial speed values of $200-400\ {\rm{km\ s}}^{-1}$ up to $450-600\ {\rm{km\ s}}^{-1}$ where a plateau is seemingly reached. The velocity error bars are computed using the $\Delta{S}$ that are assumed to be the accuracy of the HVECs' position measurements.}
\label{FIGURE_3}
\end{center}
\end{figure}

Figure~\ref{FIGURE_3}c shows the dynamical evolution of the ejected HVECs. Speeds measured at the HVECs' first detection, as estimated from the observed motion of clumps in the HI1-B images and then corrected for projection effects [i.e., $v_X/\sin\alpha$ and $v_Y/\sin\alpha$ components, where $\alpha$ is the phase angle of the comet] range between 200 and 400~km~s$^{-1}$. The ejecta undergo subsequently a large acceleration, resulting in an increase of speeds up to $450-600$~km~s$^{-1}$. The bulk speed of the ejecta seem to reach a plateau at $500-600$~km~s$^{-1}$. The latter are typical values of slow to intermediate solar wind speeds. The solid gray curve in Figure~\ref{FIGURE_3}c is a polynomial fit of all measured velocities. The dashed curves show that most of the measurements lie within the $\pm100$~km~s$^{-1}$ of the best fit. The ejecta acceleration is mainly in the X-direction [image horizontal]. Velocities in the Y-direction [image vertical] remain nearly constant with the exception of ejecta in the southern hemisphere. Velocities in the Y-direction are signed depending on the location where the clump ejection occurs [i.e., positive (negative) for clumps emitted northward (southward) of the ecliptic plane]. A dynamical equilibrium seems to be reached once velocities similar to those of the  local environment [i.e., the solar wind] are attained.

\begin{figure*}[!ht]
\begin{center}
\includegraphics[width=\textwidth]{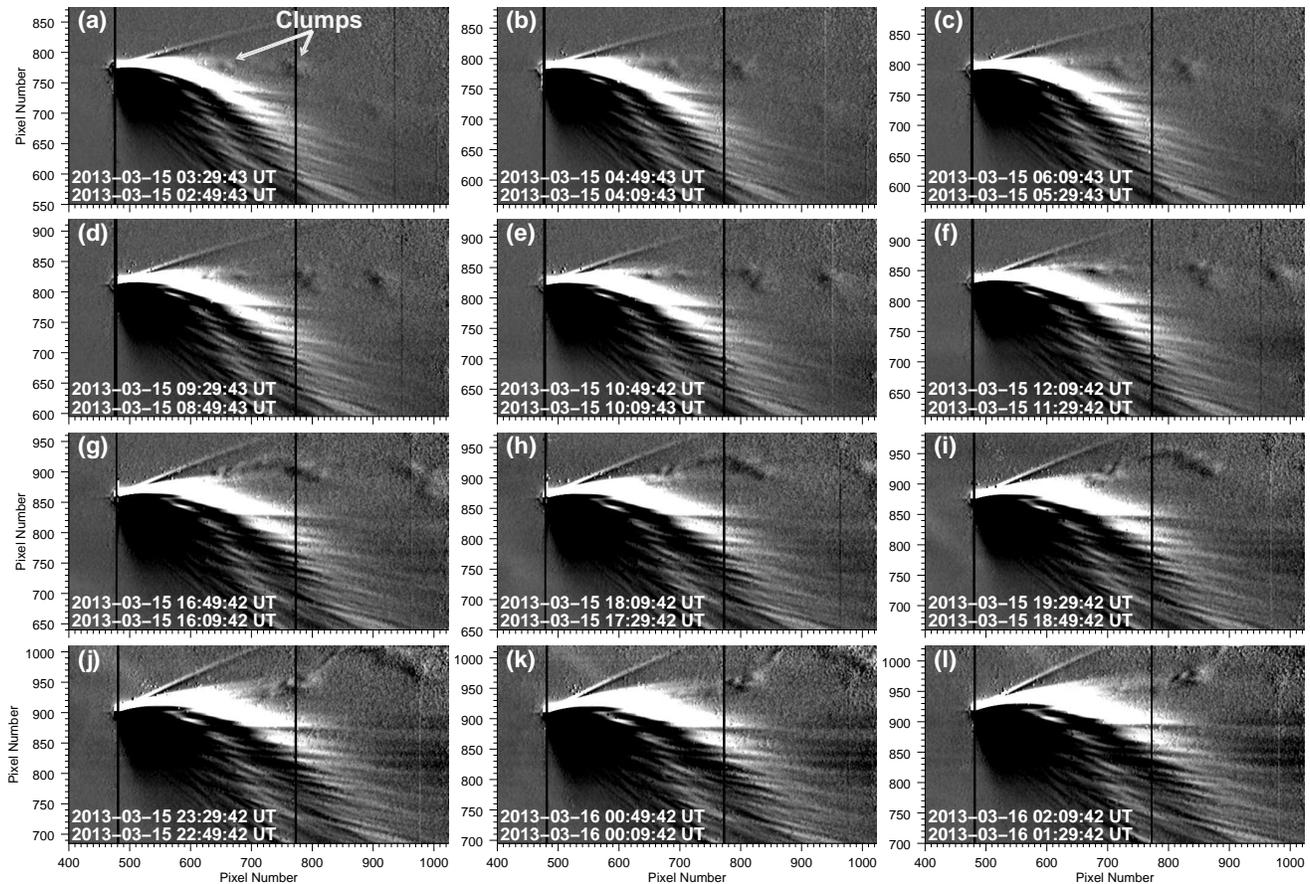}
\caption{STEREO/SECCHI/HI-1B running-difference images illustrating the HVECs ejected from the C/2011~L4 coma. The dates and times on each panel correspond to the image pair used for the running-difference images. The clumps are first detected at come- tocentric distances ranging from $\sim\!6\times10^6$~km to $\sim\!10^7$~km [$\sim\!0.041-0.076$~AU; see Table 1]. The arrows in panel (a) point to two clumps.  The HVECs initially move along near-radial structures, which progressively bend backward to form new striae in the dust tail [see panel g-l and movie in the supplementary material]. The HVECs ejection and the quasi-periodicity of the striae may correspond to distinctive activity events on the comet surface [e.g., jets].}
\label{FIGURE_2}
\end{center}
\end{figure*}

The inferred velocity values are much higher than the typical speeds of dust particles ejected from comets' nuclei \citep[i.e., $< 0.5 - 1$~km~s$^{-1}$;][]{1930PASP...42..309B,1968ApJ...154..327F,1998ApJ...496..971L,2004Icar..171..444L}. For a $\sim\!1$~$\mu{\rm{m}}$ particle, the acceleration due to solar radiation pressure is $\approx0.13$~m~s$^{-2}$ at 0.23~AU. At this acceleration, it takes $\sim\!2.7\times10^6$ seconds to reach velocities $v\sim350$~km~s$^{-1}$ in a distance of $\sim\!2\times10^8$~km [1.4 AU]. Neither of these values is in agreement with the time interval [$\sim\!2\times10^4$ sec] or distances [$\sim\!5\times10^6$ km] seen in our SECCHI observations. Moreover, unlike the Lorentz force that results in an acceleration having a $q/m$ dependence, the acceleration due to solar gravity is independent of an object's mass. This is because the acceleration due to solar radiation pressure is roughly equal to $\beta\times$ the acceleration due to solar gravity, which is also mass independent. For the Mie theory 
$$\beta=\frac{F_{{rad.\ pressure}}}{F_{{gravity}}}=\frac{0.47[\mu{\rm{m}}]}{N[{\rm{g\ cm}}^{-3}]\ R[\mu{\rm{m}}]}\approx\frac{{\rm{Surface\ Area}}}{{\rm{Volume}}};$$
where $N$ is the mass density of the dust particles and $R$ is the radius. Since the observed clumps have a $\Delta V\sim200$~km~s$^{-1}$ in about 5~hrs [Figure~\ref{FIGURE_3}c], this implies an average acceleration in that time of $\sim\!11.1$~m~s$^{-2}$, and thus an effective $\beta$ value of 11.1/0.13 = 84.6 or higher. Except for some predictions of high $\beta$ values for neutral alkali atom [due to the strong doublet  transitions in Li, Na, \& K] and neutral Ca, $\beta$ values of cometary species do not approach these values \citep[see][]{2013ApJ...771L..21F}. It is noteworthy, however, that the HI-1B transmissions for Na and K are very inefficient, only $\sim\!1\%$. C/2011~L4 would need at least $10^{25}$ to $10^{26}$ Na atoms to show up at all in HI1-B. We also note that the photoionization lifetime of these neutral atoms is on the order of $10^4$ sec \citep{2007ApJ...661L..93F}, which is long enough to stay neutral in the FOV of SECCHI. It is therefore reasonable to assume that unless some mechanism is creating large clumps of neutral alkali or Ca atoms, that solar radiation pressure does not play an important role in the acceleration of the HVECs up to $\sim200 - 400$~km~s$^{-1}$. The images [Figure~\ref{FIGURE_2}] do not show evidence for size-growth [i.e., cross-field diffusion] of the HVECs. This may suggest that the clumps may be composed of ionized species whose dynamics are controlled by the HMF.

Assuming that the acceleration is due solely to the Lorentz force $\vec{a}=\displaystyle\frac{q}{m}\vec{v}\times\vec{B}$, $v_{\rm{initial}}\approx350$~km~s$^{-1}$ [at the clump initial detection; see Figure~\ref{FIGURE_3}c], an average solar wind magnetic field strength of $|{\bf{B}}|=20-30$~nT at 0.3~AU \citep[see][]{2011P&SS...59.2075K}, and that the accelerated species are mainly composed of ionized sodium atoms [n.b. that strong Na emission lines have been reported by Wooden and Cochran [2013, priv. communication] in observations of comet C/2011~L4 from the Dunn telescope in March 2013] with $\displaystyle\frac{q}{m}=\frac{1.6\times10^{-19}}{23\times1.6\times10^{-27}}\approx4.3\times10^{6}$~C~kg$^{-1}$ results in an acceleration $a\sim\!4.5\times10^{4}$~m~s$^{-2}$. This suggests that the composition of the clumps is likely not dominated by ionized species as small as single atoms.

If on the other hand  the clumps are mainly composed of charged sub-micron dust particles with much smaller effective $q/m$ values, we can explain the observed accelerations. Specifically, if we are observing singly charged particles, then a dust particle of mass $3\times10^3$ times a Na atom's mass would exhibit the observed acceleration.  Such a particle would only have to be about 10 Na atoms in radius. Since real particles are just as likely to be made of rocky silicates than Na metal, or, following Yang et al.'s results, water ice, we should use the effective molecular weights for water ice [18 amu] or olivine [140 amu] to determine the minimum radius of the observed particles. In the case of water ice, a singly charged particle of $\sim\!11$ molecules radius [$\sim\!2\times10^{-3}\ \mu{\rm{m}}$] will exhibit the observed acceleration; for an olivine particle, a particle of only 5 [Mg$_2$SiO$_4$] units in radius [$\sim\!1.5\times10^{-3}\ \mu{\rm{m}}$] will behave as observed. These radii increase by $q_{eff}^{0.333}$, where $q_{eff}$ is the average charge on the grains. 

We thus suggest that the HVECs are most likely clouds of ionized sub-micron dust grains that are carried by the solar wind under the effect of the ${\bf{E}}\times{\bf{B}}$ forces. We cannot, however, completely rule out the possibility of neutral atoms/molecules dominated clumps.

\begin{table*}[!h]
\begin{center}
\caption{List of HVECs ejected from the C/2011~L4 coma. $r$ and PA are the heliocentric distance and position angle of the ejecta at their first appearance. PA is measured CCW from the solar north. $d_1$ and $d_2$ are the cometocentric distances of the clumps at their appearance and disappearance, respectively. $d_2^\prime$ is the disappearance cometocentric distances of the clumps with respect to the comet position at their appearance. The last column provides the variation of the comet's phase angle from the appearance to disappearance of  each HVEC. }
\begin{tabular}{c|ccccc|ccc|c|c}
\hline
 {\bf{Clump}}            & \multicolumn{5}{c|}{\bf Appearance} &  \multicolumn{3}{c|}{\bf Disappearance} & {\bf{Size}} & {\bf{Phase}}\\
{\bf{Number}}  &   Day              & Time    &  $r$    &   PA        &  $d_1$   &  Day                 &  Time  &   $d_2$ $(d_2^\prime)$  & $S\pm\Delta S$ & {\bf{Angle}}\\
  &  [03/2013]  & [UT]       &  [AU]  &   [deg.]  &  [AU]     &  [03/2013]     & [UT]& [AU]                                & ($10^5$~km) & (deg.)\\
\hline\hline
1    &  11  &  23:29  &  0.300  &  238.8  &  0.076  &  12  &  07:29  &  0.122 (0.114)  &  $7.51\pm0.83$  & 132.16 - 132.95\\
2    &  12  &  13:29  &  0.296  &  247.1  &  0.073  &  12  &  18:09  &  0.101 (0.099)  &  $8.32\pm0.26$  & 133.37 - 133.65\\
3    &  12  &  21:29  &  0.292  &  251.8  &  0.068  &  13  &  07:29  &  0.145 (0.140)  &  $9.30\pm0.55$  & 133.79 - 133.96\\
4    &  13  &  06:09  &  0.296  &  256.8  &  0.070  &  13  &  20:49  &  0.186 (0.182)  & $11.75\pm0.67$ & 133.96 - 133.64\\
5    &  13  &  23:29  &  0.305  &  267.7  &  0.070  &  14  &  15:05  &  0.171 (0.173)  &  $9.97\pm0.53$  & 133.51 - 132.30\\
6    &  14  &  20:09  &  0.308  &  279.1  &  0.059  &  15  &  12:49  &  0.188 (0.196)  & $12.50\pm0.47$ & 131.76 - 129.59\\
7    &  14  &  22:09  &  0.303  &  280.8  &  0.050  &  15  &  02:49  &  0.079 (0.081)  &  $9.42\pm0.29$  & 131.53 - 130.96\\
8    &  15  &  01:29  &  0.309  &  281.8  &  0.055  &  15  &  14:09  &  0.161 (0.167)  &  $8.12\pm0.29$  & 131.13 - 129.39\\
9    &  15  &  06:09  &  0.313  &  284.0  &  0.054  &  15  &  12:09  &  0.093 (0.096)  & $10.01\pm0.23$ & 130.52 - 129.68\\
10  &  15  &  08:49  &  0.315  &  285.4  &  0.052  &  15  &  22:49  &  0.165 (0.175)  & $12.96\pm0.51$ & 130.16 - 128.03 \\
11  &  15  &  12:09  &  0.316  &  287.6  &  0.049  &  16  &  01:29  &  0.137 (0.146)  &  $6.25\pm0.27$  & 129.68 - 127.60\\
12  &  15  &  12:49  &  0.309  &  288.7  &  0.041  &  16  &  00:09  &  0.121 (0.129)  &  $6.20\pm0.29$  & 129.59 - 127.82\\
\hline
\end{tabular}
\label{TABLE_1}
\end{center}
\end{table*}

\subsection{Clump-Striae Relationship}
Figure~\ref{FIGURE_2} and the movie provided as supplementary material show that the HVECs are expelled along bright features initially elongated near-radially [i.e., along the clumps' direction of propagation]. These elongated structures subsequently bend backward to the comet's orbital motion. They progressively evolve into new striae in the dust tail [see panels g-l of Figure~\ref{FIGURE_2}]. The HVECs-striae connection is illustrated by the movie frames $18-26$, $29-36$, $81-96$, and $109-134$ where bright clumps can be seen originating from elongated structures that result subsequently in a striae. In spite of the limited extent of the observations in time and heliodistance coverage, the data may also indicate that the timing of the bright clumps and the related, newly forming striae may originate from rotational modulation of localized areas of strong outgassing occurring on the surface of the comet's main body [i.e., nucleus] as suggested by \citet{2013ApJ...771L..21F}. This phenomena is noticeable all along the comet crossing of the HI-1B FOV and may be appreciated better through animated images [see the movie provided in the supplementary material]. A relationship between the comet nucleus' rotation and the striae spacing is reasonable as we measure a separation between the striae on the order of $10^5$~km, which at differential orbital velocities on the order of 1 to 10~km~s$^{-1}$ for the larger dust implies a time between outgassing activity maxima $\sim\!10^4$ to $\sim\!10^5$~s, or 3 to 30 hours. The comet nucleus rotation is the only regularly periodic clock in the system varying with period on the order of hours to days, and staying regular over days to weeks; the solar wind is much more stochastic on these timescales.

Although the one-to-one connection between the HVECs and striae needs to be confirmed, if this is indeed the case, it would provide new insights into the nature and formation mechanism(s) of the dust tail striae and the role of the local environment [i.e., solar wind in the present case] in the evolution and sorting of cometary dust and/or atomic/molecular species in general. It also bears clues on activity phenomena occurring on the comet surface and the dynamics of the nucleus that are not easily accessible otherwise.

\section{Discussion and Conclusions}
Detailed analysis of white-light observations from STEREO/SECCHI/HI-1B show evidence for episodical, rapidly-fading clumps ejected from the coma of C/2011~L4. They were initially observed on March 10$^{th}$, 2013, closely after C/2011~L4 entered the HI-1B FOV.  These structures are easily identified in the running-difference images initially at cometocentric distances ranging from $\sim\!6\times10^6$~km to $\sim\!10^7$~km [$\sim\!0.041-0.076\ {\rm{AU}}$; see Table~\ref{TABLE_1}] from the nucleus [see Figure~\ref{FIGURE_1} and Table~\ref{TABLE_1}]. They last for several hours and propagate to larger cometocentric distances ranging from 0.09 AU to 0.18 AU. They do not seem to undergo significant expansion and their sizes remain roughly constant. The number of ejecta increased after the comet crossed the ecliptic plane, particularly on March 14--15 [see Figure~\ref{FIGURE_2} and Table~\ref{TABLE_1}]. The bright clumps are swept near-radially in the anti-sunward direction regardless of their ejection times and locations along the comet's orbit. Their initial velocities range from 200 -- 400~km~s$^{-1}$ and are followed by an appreciable acceleration up to velocities of 450 -- 600~km~s$^{-1}$, where a plateau is seemingly reached [see Figure~\ref{FIGURE_3}]. The latter values are typical of slow and intermediate solar wind speeds. A number of these ejecta could also be traced into the HI-2B FOV.

The white-light images from HI-1B do not provide direct information about the origin of the clump material. The data indicate, however, that the HVECs may originate from discrete events occurring within the coma and are likely the result of dust and gas jets on the surface of a comet's rotating nucleus. It is unlikely that they are the result of continuous processes of dust and atomic/molecular species expulsion from the nucleus surface. \citet{2013ApJ...771L..21F} reported on spectroscopic observations showing the abundance of alkali atoms sublimated from the comet's nucleus at distances beyond the water-ice sublimation zone, providing evidence for processes energetic enough to extract atomic species and dust from the nucleus.

The STEREO image morphology indicates that dust in the coma undergoes a sorting process where very small sub-micron particles are carried anti-sunward by the solar wind while larger supra-micron particles evolve to flow into the dust tail  where they contribute to the formation of prominent striae. Whether the sub-micron particles are emitted directly from the nucleus surface, or are instead the product of a subsequent dust fragmentation process that occurs at the beginning of the striae \citep[c.f. the models of][]{1980AJ.....85.1538S,1997EM&P...78..329P,1998Icar..134...24N} is not clear - although the STEREO imagery would suggest the latter as a real possibility. Since the striae are typically only seen in very dusty comets at small heliocentric distances, we can posit the existence of a population of fractal, porous dust grains in these comets held together by materials that disrupts after a period of exposure to solar radiation and the solar wind, releasing a distribution of smaller grains that go on to form the clumps and the striae. The clump particles may also be charged throughout the interaction with local environment [i.e., solar radiation and solar wind] and are then picked up by the HMF. The dynamics of the ionized species are completely controlled by the solar wind and its frozen-in magnetic field.

Two physical processes can affect the dynamics of the dusty comet ejecta: {\bf{(I)}} Solar radiation pressure: it is expected that the material expelled from the comet nucleus lies within the sunward sector of the coma. Solar radiation pressure may contribute significantly [at least in the initial stages] to the acceleration of atomic/molecular species and sub-micron dust particles in the anti-solar direction. The influence of the solar radiation pressure on dust particles larger than $1.0$~$\mu{\rm{m}}$ is limited, which may explain the formation of new striae. If the ejected clumps are composed of dust grains, radiation pressure contributes to the initial size sorting of the dust expelled from the comet surface. {\bf{(II)}} Pick up of charged particles by the HMF: the interaction of the ejected particles with the solar wind resulting in their pick up by the local magnetic field, which may inhibit cross-field diffusion and then the expansion of the observed structures. 

Although we cannot definitively determine the composition of the ejected material, dynamics of the clump excess emission in white-light may only be attributed to (1) dusty sources via a combination of scattering by positively charged dust grains and excess free electrons due charge neutrality in the solar wind, (2) scattering by a high beta neutral atomic species, or (3) fluorescence of an ionized species. Order of magnitude calculations show that solar radiation pressure can be ruled out as the main accelerator of dust particles and that ionized single atoms or molecules accelerate too quickly compared to observations, while dust grains micron-sized or larger accelerate too slowly. Neutral Na, Li, K, or Ca atoms with $\beta>50$ could also fit the observations. The lack of expansion of the clumps may hint that these may be composed of ionized species that cannot undergo cross-field diffusion. In the case of ionized sub-micron dust material, we find that an interaction with the solar wind and the HMF can cause the observed clump dynamical evolution. We thus suggest that HVECs are clouds of ionized sub-micron dust grains that are carried by the solar wind under the effect of the ${\bf{E}}\times{\bf{B}}$ forces. We cannot, however, completely rule out the possibility of neutral atoms/molecules dominated clumps.

\begin{figure*}[!t]
\begin{center}
\includegraphics[height=0.15\textheight]{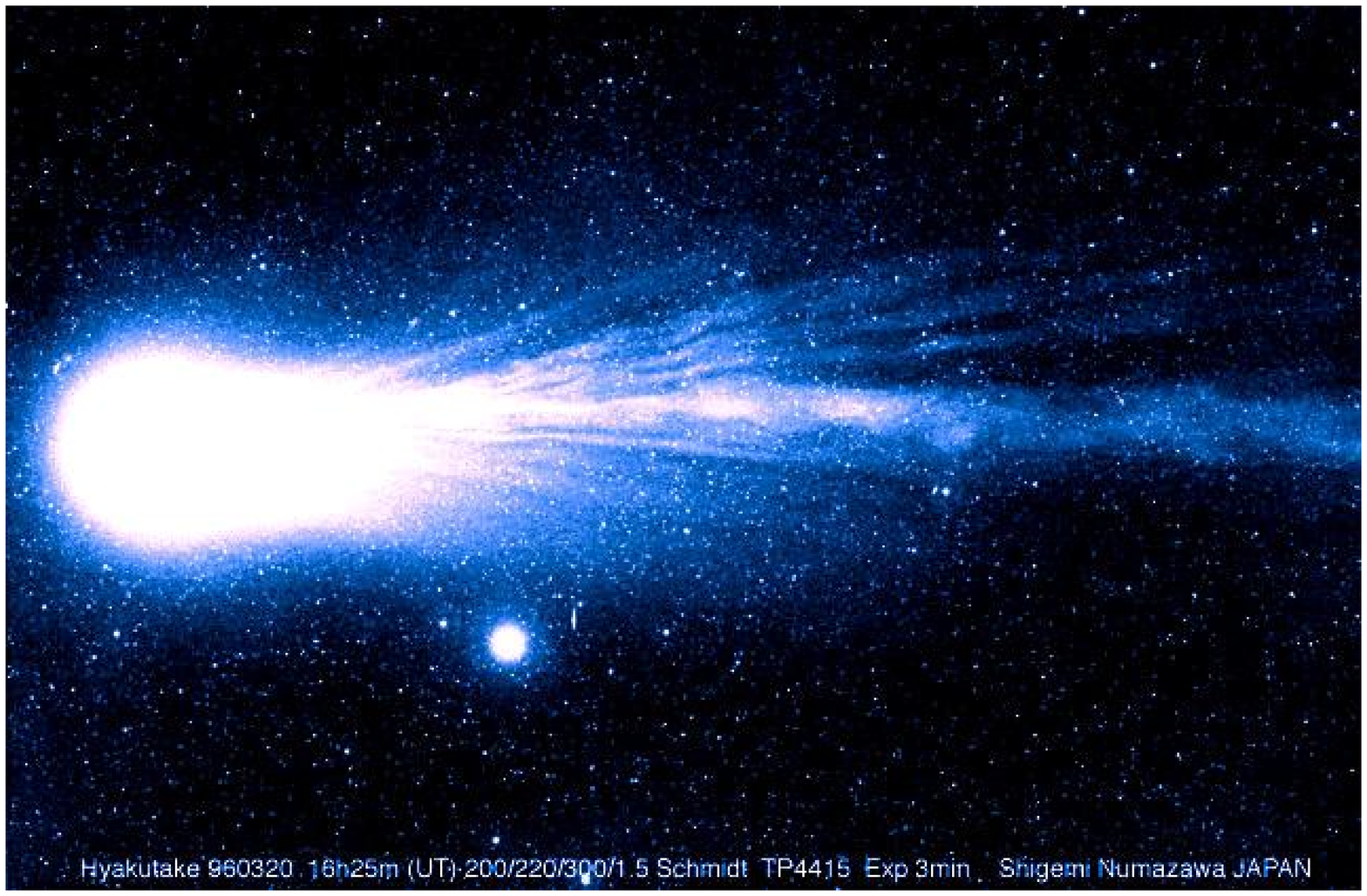}
\includegraphics[height=0.15\textheight]{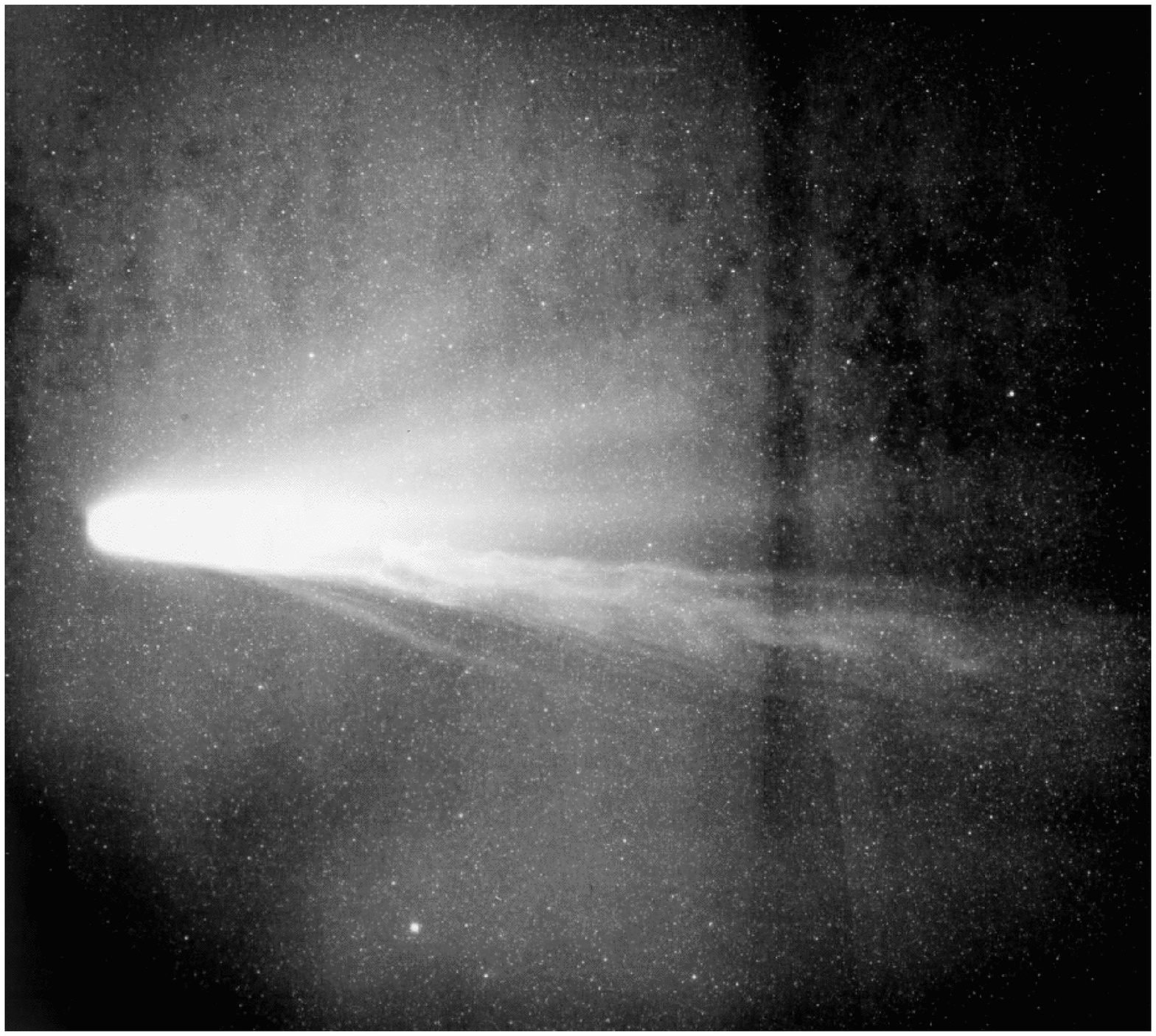}
\caption{ (Left) Image of Comet Hyakutake taken on March 20, 1996 by Shigemi Numazawa. The multiple clump-like features in the highly structured tail are striking similarity to the morphology detected for comet C/2011 L4 [Pan-STARRS] by STEREO. (Right) This morphology appeared also, although with much worse observing geometry, in Comet 1P/Halley 1986. }
\label{FIGURE_5}
\end{center}
\end{figure*}

Another interesting aspect of the HVECs is their apparent connection to newly forming striae in the dust tail. The data show that the HVECs move initially along and then detach from near-radial bright structures. The latter progressively bend backward to the comet's orbital motion and form new striae. This suggests that both phenomena may be the result of nucleus activity modulated by its rotation. The clumps-striae relationship may provide new insights into comet's nuclei activity and the role of the local environment in the evolution and sorting of cometary ejected material. The one-to-one connection between clumps and striae needs to be confirmed, however.

The nature and composition of the ejected clumps as well as the mechanisms responsible for their acceleration may be constrained by additional observations [e.g., spectroscopic] and rigorous modeling. The detailed dynamical modeling of the structures in the C/2011~L4 tail is beyond the scope of this paper, and will be studied in detail in a  follow up paper.

The STEREO observations of comet C/2011~L4 [Pan-STARRS] we have described here are compelling. But they become more interesting when we use our identification of the clump phenomenon  to search for them in other comets. The most obvious starting point for this search are in other comets known to have produced strong striae structures. Examination of the archival record shows an extremely similar coma/tail morphology for comet 1P/Halley in 1910 and 1986 [Figure~\ref{FIGURE_5}]. Clump like structures were also seen in comet C/1995~O1 [Hale-Bopp]. Not so obvious, though, were clumps in the coma/tail of comets C/2001~V1 [NEAT] as observed by SOHO or comet C/2006~P1 [McNaught] as observed from earth, despite the latter comet displaying an extraordinary sweep of striae across many degrees of the evening twilight sky. Future models of clumps and striae will have to rate this variability into account. On the other hand, the dust and gas composition, dust size distribution and dust/gas emission rates are known to vary significantly from comet to comet, so it may be the variation in one of these quantities [e.g., the sodium atom abundance] that controls the clump brightness.

\begin{acknowledgments}
The authors are grateful for the anonymous referees for constructive comments and suggestions. We would like to thank A. Vourlidas [JHUAPL] for helpful comments and suggestions. NER would like to thank NASA's Solar Probe Plus mission for supporting partially the present work. GHJ is partially supported by the Science and Technology Facilities Council, UK. The data used in the present analysis were acquired from the Virtual Solar Observatory [VSO: http://www.virtualsolar.org] database and were processed to higher levels using the STEREO/SECCHI/HI-1 calibration procedures available on Solar Software tree [SSW: http://www.lmsal.com/solarsoft]. The STEREO/SECCHI data are produced by an international consortium of the NRL [USA], LMSAL [USA], NASA-GSFC [USA], RAL [UK], University of Birmingham [UK], MPS [Germany], CSL [Belgium], IOTA [France], and IAS [France].
\end{acknowledgments}



\end{article}

\begin{thebibliography}{}
\providecommand{\natexlab}[1]{#1}
\expandafter\ifx\csname urlstyle\endcsname\relax
  \providecommand{\doi}[1]{doi:\discretionary{}{}{}#1}\else
  \providecommand{\doi}{doi:\discretionary{}{}{}\begingroup
  \urlstyle{rm}\Url}\fi

\bibitem[A'Hearn et al.(2005)]{2005Sci...310..258A} A'Hearn, M. F., M. J. S., Belton, W. A., Delamere, J., Kissel, K. P., Klaasen, L. A., McFadden, K. J., Meech, H. J., Melosh, P. H., Schultz, J. M., Sunshine, P. C., Thomas, J., Veverka, D. K., Yeomans, M. W., Baca, I., Busko, C. J., Crockett, S. M., Collins, M., Desnoyer, C. A., Eberhardy, C. M., Ernst, T. L., Farnham, L., Feaga, O., Groussin, D., Hampton, S. I., Ipatov, J.-Y., Li, D., Lindler, C. M., Lisse, N., Mastrodemos, W. M., Owen, J. E., Richardson, D. D., Wellnitz, \& R. L., White, 2005, ``Deep Impact: Excavating Comet Tempel 1", Science, 310, 258, \doi{10.1126/science.1118923}.

\bibitem[Bar-Nun \& Laufer(2003)]{2003Icar..161..157B} Bar-Nun, A., \& D., Laufer, 2003, ``First experimental studies of large samples of gas-laden amorphous ``cometary'' ices", Icarus, 161, 157, \doi{10.1016/S0019-1035(02)00016-7}.

\bibitem[Bewsher et al.(2012)]{2012SoPh..276..491B} Bewsher, D., D.~S., Brown, \& C.~J., Eyles, 2012, ``Long-Term Evolution of the Photometric Calibration of the STEREO Heliospheric Imagers: I. HI-1", Sol. Phys., 276, 491, \doi{10.1007/s11207-011-9874-7}.


\bibitem[Bobrovnikoff(1930)]{1930PASP...42..309B} Bobrovnikoff, N.~T., 1930, ``Halley's Comet in 1910", \pasp 42, 309, \doi{10.1086/124062}.

\bibitem[Boehnhardt(2004)]{2004come.book..301B} Boehnhardt, H., 2004, ``Split comets", IN: Comets II, M. C. Festou, H. U. Keller, and H. A. Weaver (eds.), University of Arizona Press, Tucson, 745, pp.301-316, Bibliographic Code: 2004come.book..301B. 

\bibitem[Brandt(1968)]{1968ARA&A...6..267B} Brandt, J.~C., 1968, ``The Physics of Comet Tails", ARA\&A, 6, 267, \doi{10.1146/annurev.aa.06.090168.001411}.

\bibitem[Brownlee et al.(2004)]{2004Sci...304.1764B} Brownlee, D. E., F., Horz, R. L., Newburn, M., Zolensky, T. C., Duxbury, S., Sandford, Z., Sekanina, P., Tsou, M. S., Hanner, B. C., Clark, S. F., Green, \& J., Kissel, 2004, ``Surface of Young Jupiter Family Comet 81 P/Wild 2: View from the Stardust Spacecraft", Science, 304, 1764, \doi{10.1126/science.1097899}.

\bibitem[Buffington et al.(2008)]{2008ApJ...677..798B} Buffington, A., M. M., Bisi, J. M., Clover, P. P., Hick, B. V., Jackson, \& T. A., Kuchar, 2008, ``Analysis of Plasma-Tail Motions for Comets C/2001 Q4 (NEAT) and C/2002 T7 (LINEAR) Using Observations from SMEI", \apj 677, 798, \doi{10.1086/529039}.

\bibitem[Clark et al.(2004)]{2004JGRE..10912S03C} Clark, B. C., S. F., Green, T. E., Economou, S. A., Sandford, M. E., Zolensky, N., McBride, \& D. E., Brownlee, 2004, ``Release and fragmentation of aggregates to produce heterogeneous, lumpy coma streams", \jgr (Planets), 109, E12S03, \doi{10.1029/2004JE002319}.

\bibitem[Clover et al.(2010)]{2010ApJ...713..394C} Clover, J.~M., B.~V., Jackson, A., Buffington, P.~P., Hick, \& M.~M. Bisi, 2010, ``Solar Wind Speed Inferred from Cometary Plasma Tails using Observations from STEREO HI-1", \apj 713, 394, \doi{10.1088/0004-637X/713/1/394}.

\bibitem[Combi et al.(2014)]{2014AJ....147..126C} Combi, M. R., J.-L., Bertaux, E., Qu\'emerais, S., Ferron, J. T. T., M\"akinen, \& G., Aptekar, 2014, ``Water Production in Comets C/2011 l4 (PanSTARRS) and C/2012 f6 (Lemmon) from observations with SOHO/SWAN", \aj 147, 126, \doi{10.1088/0004-6256/147/6/126}.

\bibitem[Ershkovich(1976)]{1976P&SS...24..287E} Ershkovich, A.~I., 1976, ``Solar wind interaction with the tail of Comet Kohoutek", Planet. Space Sci., 24, 287, \doi{10.1016/0032-0633(76)90025-8}.

\bibitem[Eyles et al.(2009)]{2009SoPh..254..387E} Eyles, C. J., R. A., Harrison, C. J., Davis, N. R., Waltham, B. M., Shaughnessy, H. C. A., Mapson-Menard, D., Bewsher, S. R., Crothers, J. A., Davies, G. M., Simnett, R. A., Howard, J. D., Moses, J. S., Newmark, D. G., Socker, J.-P., Halain, J.-M., Defise, E., Mazy \& P., Rochus, 2009, ``The Heliospheric Imagers Onboard the STEREO Mission", Sol. Phys., 254, 387, \doi{10.1007/s11207-008-9299-0}.

\bibitem[Farnham \& Meech(1994)]{1994ApJS...91..419F} Farnham, T.~L., \& K.~J., Meech, 1994, ``Comparison of the plasma tails of four comets: P/Halley, Okazaki-Levy-Rudenko, Austin, and Levy", \apjs 91, 419, \doi{10.1086/191943}.

\bibitem[Finson \& Probstein(1968)]{1968ApJ...154..327F} Finson, M.~J., \& R.~F., Probstein, 1968, ``A theory of dust comets. I. Model and equations", \apj 154, 327, \doi{10.1086/149761}.

\bibitem[Fulle et al.(2007)]{2007ApJ...661L..93F} Fulle, M., F., Leblanc, R. A., Harrison, C. J., Davis, C. J., Eyles, J. P., Halain, R. A., Howard, D., Bockel\'ee-Morvan, G., Cremonese, \& T., Scarmato, 2007, ``Discovery of the Atomic Iron Tail of Comet MCNaught Using the Heliospheric Imager on STEREO", \apjl, 661, L93, \doi{10.1086/518719}.

\bibitem[Fulle et al.(2013)]{2013ApJ...771L..21F} Fulle, M., P., Molaro, L., Buzzi, \& P., Valisa, 2013, ``Potassium Detection and Lithium Depletion in Comets C/2011 L4 (Panstarrs) and C/1965 S1 (Ikeya-Seki)", \apj 771, L21, \doi{10.1088/2041-8205/771/2/L21}.

\bibitem[Glassmeier et al.(2007)]{2007SSRv..128....1G} Glassmeier, K.-H., H., Boehnhardt, D., Koschny, E., K{\"u}hrt, \& I., Richter, 2007, ``The Rosetta Mission: Flying Towards the Origin of the Solar System", Space Sci. Rev., 128, 1, \doi{10.1007/s11214-006-9140-8}.

\bibitem[Hill \& Mendis(1980)]{1980ApJ...242..395H} Hill, J.~R., \& D.~A., Mendis, 1980, ``On the origin of striae in cometary dust tails", \apj 242, 395, \doi{10.1086/158472}.

\bibitem[Howard et al.(2008)]{2008SSRv..136...67H} Howard, R. A., J. D., Moses, A., Vourlidas, J. S., Newmark, D. G., Socker, S. P., Plunkett, C. M., Korendyke, J. W., Cook, A., Hurley, J. M., Davila, W. T., Thompson, O. C., St Cyr, E., Mentzell, K., Mehalick, J. R., Lemen, J. P., Wuelser, D. W., Duncan, T. D., Tarbell, C. J., Wolfson, A., Moore, R. A., Harrison, N. R., Waltham, J., Lang, C. J., Davis, C. J., Eyles, H., Mapson-Menard, G. M., Simnett, J. P., Halain, J. M., Defise, E., Mazy, P., Rochus, R., Mercier, M. F., Ravet, F., Delmotte, F., Auch\`ere, J. P., Delaboudini\`ere, V., Bothmer, W., Deutsch, D., Wang, N., Rich, S., Cooper, V., Stephens, G., Maahs, R., Baugh, D., McMullin, \& T., Carter, 2008, ``Sun Earth Connection Coronal and Heliospheric Investigation (SECCHI)", Space Sci. Rev., 136, 67, \doi{10.1007/s11214-008-9341-4}.

\bibitem[Jackson et al.(2013)]{2013AIPC.1539..364J} Jackson, B. V., A., Buffington, J. M., Clover, P. P., Hick, H.-S., Yu, \& M. M., Bisi, 2013, ``Using comet plasma tails to study the solar wind", American Institute of Physics Conference Series, 1539, 364, \doi{10.1063/1.4811062}.

\bibitem[Jia et al.(2009)]{2009ApJ...696L..56J} Jia, Y. D., C. T., Russell, L. K., Jian, W. B., Manchester, O., Cohen, A., Vourlidas, K. C., Hansen, M. R., Combi, \& T. I., Gombosi, 2009, ``Study of the 2007 April 20 CME-Comet Interaction Event with an MHD Model", \apjl 696, L56, \doi{10.1088/0004-637X/696/1/L56}.

\bibitem[Jones \& Brandt(2004)]{2004GeoRL..3120805J} Jones, G.~H., \& J.~C., Brandt, 2004, ``The interaction of comet 153P/Ikeya-Zhang with interplanetary coronal mass ejections: Identification of fast ICME signatures", \grl 31, L20805, \doi{10.1029/2004GL021166}.

\bibitem[Jones et al.(2004)]{2004DPS....36.2102J} Jones, G. H., J. S., Morrill, D., Hammer, C. M., Lisse, T. L., Farnham, \& G. R., Lawrence, 2004, ``Comet C/2002 V1 (NEAT) - Evidence of solar wind effects on a comet's ion and dust tails at 0.1 AU", American Astronomical Society, DPS meeting \#36, \#21.02, Bulletin of the American Astronomical Society, Vol. 36, p.1117, Bibliographic Code: 2004DPS....36.2102J.

\bibitem[Kaiser et al.(2008)]{2008SSRv..136....5K} Kaiser, M. L., T. A., Kucera, J. M., Davila, O. C., St. Cyr, M., Guhathakurta, \& E., Christian, 2008, ``The STEREO Mission: An Introduction", Space Sci. Rev., 136, 5, \doi{10.1007/s11214-007-9277-0}.

\bibitem[Keller et al.(1986)]{1986Natur.321..320K} Keller, H. U., C., Arpigny, C., Barbieri, R. M., Bonnet, S., Cazes, M., Coradini, C. B., Cosmovici, W. A., Delamere, W. F., Huebner, D. W., Hughes, C., Jamar, D., Malaise, H. J., Reitsema, H. U., Schmidt, W. K. H., Schmidt, P., Seige, F. L., Whipple, \& K., Wilhelm, 1986, ``First Halley multicolour camera imaging results from Giotto", Nature, 321, 320, \doi{10.1038/321320a0}.

\bibitem[Keller et al.(1987)]{1987A&A...187..807K} Keller, H.~U., W.~A., Delamere, H.~J., Reitsema, W.~F., Huebner, \& H.~U., Schmidt, 1987, ``Comet P/Halley's nucleus and its activity", \aap 187, 807, Bibliographic Code: 1987A\&A...187..807K.

\bibitem[Kimura et al.(2002)]{2002Icar..157..349K} Kimura, H., H., Okamoto, \& T., Mukai, 2002, ``Radiation Pressure and the Poynting-Robertson Effect for Fluffy Dust Particles", Icarus, 157, 349, \doi{10.1006/icar.2002.6849}.

\bibitem[Konno et al.(1990)]{1990LPI....21..653K} Konno, I., W.~F., Huebner, \& D.~C. Boice,, 1990, ``A Hydrodynamic Model of Dusty Gas Flows in Comet Comae with Dust Fragmentation", Lunar and Planetary Science Conference, 21, 653, Bibliographic Code:1990LPI....21..653K.

\bibitem[Konno et al.(1993)]{1993Icar..101...84K} Konno, I., W.~F., Huebner, \& D.~C. Boice,, 1993, ``A model of dust fragmentation in near-nucleus jet-like features on Comet P/Halley", Icarus, 101, 84, \doi{10.1006/icar.1993.1008}.

\bibitem[Korth et al.(2011)]{2011P&SS...59.2075K} Korth, H., B. J., Anderson, T. H., Zurbuchen, J. A., Slavin, S., Perri, S. A., Boardsen, D. N., Baker, S. C., Solomon, \& R. L., McNutt, 2011, ``The interplanetary magnetic field environment at Mercury's orbit", Planet. Space Sci., 59, 2075, \doi{10.1016/j.pss.2010.10.014}.

\bibitem[Kramer et al.(2014)]{2014Icar..236..136K} Kramer, E.~A., Y.~R., Fernandez, C.~M., Lisse, M.~S.~P., Kelley, \& L.~M., Woodney, 2014, ``A dynamical analysis of the dust tail of Comet C/1995 O1 (Hale-Bopp) at high heliocentric distances", Icarus, 236, 136, \doi{10.1016/j.icarus.2014.03.033}.

\bibitem[Kr{\'o}likowska et al.(2012)]{2012A&A...544A.119K} Kr{\'o}likowska, M., P.~A., Dybczy{\'n}ski, \& G., Sitarski, 2012, ``Different dynamical histories for comets C/2001 Q4 and C/2002 T7?", \aap 544, AA119, \doi{10.1051/0004-6361/201219408}.

\bibitem[Lee et al.(2014)]{2014Icar..239..141L} Lee, S., M., Hofstadter, M. A., Frerking, S., Gulkis, P., von Allmen, J., Crovisier, N., Biver, D., Bockel\'ee-Morvan, L., Kamp, M., Choukroun, S., Keihm, \& M., Janssen, 2014, ``Sub-millimeter observation of water vapor at 557 GHz in Comet C/2002 T7 (LINEAR)", Icarus, 239, 141, \doi{10.1016/j.icarus.2014.05.004}.

\bibitem[Lisse et al.(1998)]{1998ApJ...496..971L} Lisse, C. M., M. F., A'Hearn, M. G., Hauser, T., Kelsall, D. J., Lien, S. H., Moseley, W. T., Reach, \& R. F., Silverberg, 1998, ``Infrared Observations of Comets by COBE", \apj 496, 971, \doi{10.1086/305397}.

\bibitem[Lisse et al.(2004)]{2004Icar..171..444L} Lisse, C. M., Y. R., Fern\'andez, M. F., A'Hearn, E., Gr\"un, H. U., K\"aufl, D. J., Osip, D. J., Lien, T., Kostiuk, S. B., Peschke, \& R. G., Walker, 2004, ``A tale of two very different comets: ISO and MSX measurements of dust emission from 126P/IRAS (1996) and 2P/Encke (1997)", Icarus, 171, 444, \doi{10.1016/j.icarus.2004.05.015}.

\bibitem[Meech \& Svoren(2004)]{2004come.book..317M} Meech, K.~J., \& J., Svoren, 2004, ``Using cometary activity to trace the physical and chemical evolution of cometary nuclei", IN: Comets II, M. C. Festou, H. U. Keller, and H. A. Weaver (eds.), University of Arizona Press, Tucson, 745 pp., p.317-335, Bibliographic Code: 2004come.book..317M.

\bibitem[Mendis et al.(1986)]{1986GeoRL..13..239M} Mendis, D. A., E. J., Smith, B. T., Tsurutani, J. A., Slavin, D. E., Jones, \& G. L., Siscoe, 1986, ``Comet-solar wind interaction - Dynamical length scales and models", \grl 13, 239, \doi{10.1029/GL013i003p00239}.

\bibitem[Mendis \& Hor{\'a}nyi(1991)]{1991GMS....61...17M} Mendis, D.~A., \& M., Hor{\'a}nyi, 1991, ``Dust-plasma interactions in the cometary environment", IN: Cometary plasma processes (A92-10001 01-90). Washington, DC, American Geophysical Union, p. 17-25, Bibliographic Code: 1991GMS....61...17M.

\bibitem[Mendis \& Hor{\'a}nyi(2013)]{2013RvGeo..51...53M} Mendis, D.~A., \& M., Hor{\'a}nyi, 2013, ``Dusty Plasma Effects in Comets: Expectations for Rosetta", Rev. Geophys., 51, 53, \doi{10.1002/rog.20005}.

\bibitem[Nishioka \& Watanabe(1990)]{1990Icar...87..403N} Nishioka, K., \& J.-I., Watanabe, 1990, ``Finite lifetime fragment model for synchronic band formation in dust tails of comets", Icarus, 87, 403, \doi{10.1016/0019-1035(90)90143-W}.

\bibitem[Nishioka(1998)]{1998Icar..134...24N} Nishioka, K., 1998, ``Finite Lifetime Fragment Model 2 for Synchronic Band Formation in Dust Tails of Comets", Icarus, 134, 24, \doi{10.1006/icar.1998.5935}.

\bibitem[Notesco et al.(2003)]{2003Icar..162..183N} Notesco, G., A., Bar-Nun, \& T., Owen, 2003, ``Gas trapping in water ice at very low deposition rates and implications for comets", Icarus, 162, 183, \doi{10.1016/S0019-1035(02)00059-3}.

\bibitem[Notni \& Thaenert(1988)]{1988AN....309..133N} Notni, P., \& W., Thaenert, 1988, ``The striae in the dust tails of great comets - A comparison to various theories", Astronomische Nachrichten, 309, 133, Bibliographic Code: 1988AN....309..133N.

\bibitem[Oppenheim et al.(1968)]{OppenheimEtAl1968} Oppenheim, A.V., R.W., Schafer, \& T.G., Jr., Stockham, 1968, ``Nonlinear filtering of multiplied and convolved signals", Proc. IEEE, {\it{Vol.}} 56(8), pp.1264-1291, \doi{10.1109/TAU.1968.1161990}.


\bibitem[Prialnik(1992)]{1992ApJ...388..196P} Prialnik, D., 1992, ``Crystallization, sublimation, and gas release in the interior of a porous comet nucleus", \apj 388, 196, \doi{10.1086/171143}.

\bibitem[Pittichov{\'a} et al.(1997)]{1997EM&P...78..329P} Pittichov{\'a}, J., Z., Sekanina, K., Birkle, H., Boehnhardt, D., Engels, \& P., Keller, 1997, ``An Early Investigation Of The Striated Tail Of Comet Hale-Bopp (C/1995 O1)", Earth, Moon, and Planets, 78, 329, \doi{10.1023/A:1006242209416}.

\bibitem[Russell et al.(1982)]{1982come.coll..561R} Russell, C.~T., J.~G., Luhmann, R.~C., Elphic, \& M., Neugebauer, 1982, ``Solar wind interaction with comets - Lessons from Venus", IAU Colloq.~61: Comet Discoveries, Statistics, and Observational Selection, p.561-587, Bibliographic Code: 1982come.coll..561R.

\bibitem[Sekanina \& Farrell(1980)]{1980AJ.....85.1538S} Sekanina, Z., \& J.~A., Farrell, 1980, ``The striated dust tail of Comet West 1976 VI as a particle fragmentation phenomenon", \aj 85, 1538, \doi{10.1086/112831}.

\bibitem[Sekanina \& Farrell(1982)]{1982AJ.....87.1836S} Sekanina, Z., \& J.~A., Farrell, 1982, ``Two dust populations of particle fragments in the striated tail of Comet MRKOS 1957 V", \aj 87, 1836, \doi{10.1086/113274}.

\bibitem[Sekanina et al.(2004)]{2004Sci...304.1769S} Sekanina, Z., D.~E., Brownlee, T.~E., Economou, A.~J., Tuzzolino, \& S.~F., Green, 2004, ``Modeling the Nucleus and Jets of Comet 81P/Wild 2 Based on the Stardust Encounter Data", Science, 304, 1769, \doi{10.1126/science.1098388}.

\bibitem[Sekanina(2007)]{2007SPIE.6694E..0IS} Sekanina, Z., 2007, ``Dust jets, outbursts, and fragmentation of comets", IN: Instruments, Methods, and Missions for Astrobiology X. Edited by Hoover, Richard B., Levin, Gilbert V., Rozanov, Alexei Y., Davies, Paul C. W.. Proceedings of the SPIE, Volume 6694, article id. 66940I, \doi{10.1117/12.732612}.

\bibitem[Socker et al.(2000)]{2000SPIE.4139..284S} Socker, D.~G., R.~A., Howard, C.~M., Korendyke, G.~M., Simnett, \& D.~F., Webb, 2000, ``NASA Solar Terrestrial Relations Observatory (STEREO) mission heliospheric imager", Proc. SPIE Vol. 4139, p. 284-293, Instrumentation for UV/EUV Astronomy and Solar Missions, S. Fineschi, C. M. Korendyke, O. H. Siegmund, B. E. Woodgate, Eds, Bibliographic Code: 2000SPIE.4139..284S.

\bibitem[Soderblom et al.(2002a)]{2002Sci...296.1087S} Soderblom, L. A., T. L., Becker, G., Bennett, D. C., Boice, D. T., Britt, R. H., Brown, B. J., Buratti, C., Isbell, B., Giese, T., Hare, M. D., Hicks, E., Howington-Kraus, R. L., Kirk, M., Lee, R. M., Nelson, J., Oberst, T. C., Owen, M. D., Rayman, B. R., Sandel, S. A., Stern, N., Thomas, \& R. V., Yelle, 2002a, ``Observations of Comet 19P/Borrelly by the Miniature Integrated Camera and Spectrometer Aboard Deep Space 1", Science, 296, 1087, \doi{10.1126/science.1069527}.

\bibitem[Soderblom et al.(2002b)]{2002LPI....33.1256S} Soderblom, L. A., T. L., Becker, G., Bennett, D. C., Boice, D. T., Britt, R. H., Brown, B. J., Buratti, C., Isbell, B., Giese, T., Hare, M. D., Hicks, E., Howington-Kraus, R. L., Kirk, M., Lee, R. M., Nelson, J., Oberst, T., Owen, B. R., Sandel, S. A., Stern, N., Thomas, R. V., Yelle, 2002b, ``Encounter with Comet 19P/Borrelly: Results from the Deep Space 1 Miniature Integrated Camera and Spectrometer", Lunar and Planetary Science Conference, 33, 1256, Bibliographic Code: 2002LPI....33.1256S.

\bibitem[Tuzzolino et al.(2004)]{2004Sci...304.1776T} Tuzzolino, A. J., T. E., Economou, Ben C., Clark, P., Tsou, D. E., Brownlee, S. F., Green, J. A. M., McDonnell, N., McBride, \& M. T. S. H., Colwell, 2004, ``Dust Measurements in the Coma of Comet 81P/Wild 2 by the Dust Flux Monitor Instrument", Science, 304, 1776, \doi{10.1126/science.1098759}.

\bibitem[Wainscoat et al.(2011)]{2011IAUC.9215....1W} Wainscoat, R., M., Micheli, L. ,Wells, R., Holmes, S., Foglia, T., Vorobjov, G., Sostero, E., Guido, H., Sato, \& G. V., Williams, 2011, ``Comet C/2011 L4 (Panstarrs)", IAU circ., 9215, 1, Bibliographic Code: 2011IAUC.9215....1W.

\bibitem[Whipple(1951)]{1951ApJ...113..464W} Whipple, F.~L., 1951, ``A Comet Model. II. Physical Relations for Comets and Meteors", \apj 113, 464, \doi{10.1086/145416}.

\bibitem[Yang et al.(2014)]{2014ApJ...784L..23Y} Yang, B., J., Keane, K., Meech, T., Owen, \& R., Wainscoat, 2014, ``Multi-wavelength Observations of Comet C/2011 L4 (Pan-STARRS)", \apj 784, L23, \doi{10.1088/2041-8205/784/2/L23}.

\bibitem[Yelle et al.(2004)]{2004Icar..167...30Y} Yelle, R.~V., L.~A., Soderblom, \& J.~R., Jokipii, 2004, ``Formation of jets in Comet 19P/Borrelly by subsurface geysers", Icarus, 167, 30, \doi{10.1016/j.icarus.2003.08.020}.

\end{thebibliography}
\end{document}